\documentclass[fleqn,twoside]{article}
\usepackage{espcrc2}
\usepackage{epsfig}

\usepackage{graphicx}
\usepackage[figuresright]{rotating}

\newcommand{\AmS}{{\protect\the\textfont2
  A\kern-.1667em\lower.5ex\hbox{M}\kern-.125emS}}
\newcommand{\be}{\begin{equation}}
\newcommand{\ee}{\end{equation}}
\newcommand{\ba}{\begin{eqnarray}}
\newcommand{\ea}{\end{eqnarray}}
\newcommand{\tr}{{\rm Tr\,}}
\newcommand{\ii}{{\rm i}}

\newcommand{\ex}{{\rm e}}

\newcommand{\bfx}{{\bf x}}

\newcommand{\eq}{Eq.~}
\newcommand{\eqs}{Eqs.~}
\newcommand{\fig}{Fig.~}
\title{Non-perturbative parton mass for the gluon
       }

\author{O. Philipsen\address[CTP]{Center for Theoretical Physics, 
        Massachusetts Institute of Technology, \\ 
        Cambridge, MA 02139, USA}}
\begin{document}

\begin{abstract}
A gauge invariant, non-local observable is constructed in pure gauge theory,
which is identical to the gluon propagator in a particular gauge, permitting to define
a non-perturbative parton mass for the gluon. This mass can be shown to be
related to the $1P-1S$ mass splitting of heavy quarkonia. Preliminary numerical results 
for 3d SU(2) yield $m_A=0.37(6)g^2$, while from the $\bar{b}b$ spectrum one infers 
$m_A\approx 420$ MeV for QCD.
\vspace{1pc}
\end{abstract}

\maketitle

\section{INTRODUCTION}

Knowledge of non-perturbative parton properties is essential to
understand the interplay between
high energy parton physics and low energy hadron physics.
At finite temperature the deconfinement transition
interpolates between the two regimes. The 
high temperature phase is expected to be governed by parton dynamics,
which is encoded in the Green functions of
quarks and gluons. In general those are not gauge invariant.

In perturbation theory, one fixes a gauge and studies partons directly.
Although field propagators are not physical observables, gauge invariant 
dynamical information is carried by their singularity structure.
For example,
pole masses defined from the gluon and quark propagators have been proved to
be gauge invariant order by order in perturbation theory \cite{kro}. 
Resummation
schemes have been designed to self-consistently compute the pole of the gluon propagator 
in three dimensions \cite{mm}, which is related to the ``magnetic mass'' of the thermal 
gauge theory. However,
it has remained an open question whether a pole exists non-perturbatively.

On the other hand, 
numerical gauge fixing on the lattice is problematic:
It is difficult to avoid Gribov copies and fix a gauge uniquely.
Most complete gauge fixings (e.g. the Landau gauge)
violate the positivity of the transfer matrix, thus obstructing
a quantum mechanical interpretation of the results.
For quark mass computations, gauge fixing can be avoided by non-perturbative renormalization techniques. 
However, those rely on hadronic quantities through PCAC relations and are not
applicable to gluons. A way around these problems has been suggested \cite{mm}
by employing spatially non-local, composite operators. Without loss of generality, SU(2)
pure gauge theory will be considered in the following. 

\section{A NON-LOCAL GLUON OPERATOR}

A gauge invariant gluon operator can be defined when a complex $N$-plet
transforming in the fundamental representation is available.
One possibility is to take the eigenfunctions
of the spatial covariant Laplacian, which is a hermitian operator with a positive spectrum,
\be \label{lev}
-\left(D_i^2[U]\right)_{\alpha\beta}f^{(n)}_\beta(x)=
%\sum_{i=1,d}\left[2f(x)-U_i(x)f^{(n)}(x+\hat{\i})-U^\dag_i(x-\hat{\i})
%f^{(n)}(x-\hat{\i})\right]=
\lambda_n f^{(n)}_\alpha(x),  \quad \lambda^n>0.
\ee
They provide a unique mapping $U\rightarrow f[U]$ except when eigenvalues are degenerate
or $|f|=0$. In simulations the probability of generating such
configurations is essentially zero.
These properties have been used previously for gauge fixing
without Gribov copies and to construct blockspins for the derivation of
effective theories \cite{lap}. 

\begin{figure}[th]
%\vspace*{-0.1cm}
\centerline{\epsfxsize=5cm\hspace*{0cm}\epsfbox{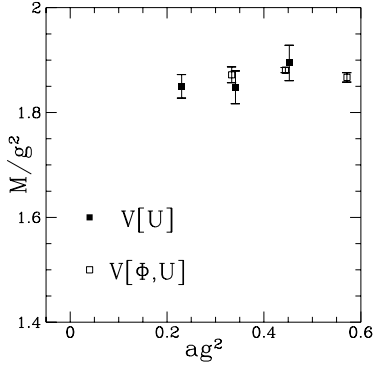}}
\vspace*{-1cm}
\caption[]{{
The W-boson mass in a 3d Higgs phase, computed from the standard operator
$V[\phi,U]$ and the non-local $V[U]$, \eqs (\ref{cl}).}}
\label{comp}
\end{figure}

The lowest eigenvectors are used to construct the matrix
$\Omega(x)\equiv\frac{1}{|f(x)|}\left(
f^{(1)}(x),f^{(2)}(x)\right)
$, which transforms as $\Omega^g(x)=g(x)\Omega(x)h^\dag (t)$.
We can now define composite link and gluon fields
\ba \label{cl}
V_\mu(x)&=&\Omega^\dag(x)U_\mu(x)\Omega(x+\hat{\mu}),\\
A_\mu(x)&=&\frac{\ii}{2g}\left(V_\mu(x)-V^\dag_\mu(x)\right),
\ea
both transforming as
$O_i^g(x)=h(t)O_i(x)h^\dag(t)$, whereas $V_0^g(x)=h(t)V_0(x)h^\dag(t+1)$.
The zero momentum projected time links 
$\tilde{V_0}(t)=\sum_{\bfx}V_0(\bfx,t)/|\sum_{\bfx}V_0(\bfx,t)|$
are multiplied to ``strings'' $\tilde{V}_0(t_1,t_2)$ connecting two timeslices.
These ingredients can be combined to the gauge invariant operator
\be \label{ofinal}
O[U]=\tr \left[A_i(\bfx,0)\tilde{V}_0(0,t)
A_i(\bfx,t)\tilde{V}_0^\dag(0,t)\right],
\ee
which in the particular gauge $V_0(t)=1$ reduces to the gluon propagator.
In \cite{me} the transfer matrix formalism was used to show that \eq (\ref{ofinal})
falls off exponentially as
\be \label{finalen}
\sim \sum_{n}
|\langle 0|\hat{A}_{\alpha\beta}(\bfx)
|n^{L0}\rangle |^2 \ex^{ -(E_n-E_0)t}\;.
\ee
The eigenvalues $E_n$ and eigenvectors $|n^{L0}\rangle$ are those of the
Hamiltonian in Laplacian temporal gauge, $V_0(x)=1$.
It has been proved \cite{me} that this Hamiltonian has the same
spectrum as the Kogut-Susskind Hamiltonian, which is obtained by quantizing
the theory in temporal gauge \cite{ks}.

\begin{figure}[th]
\centerline{\epsfxsize=6cm\hspace*{0cm}\epsfbox{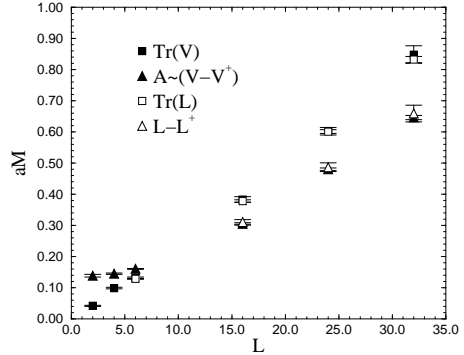}}
\vspace*{-1.2cm}
\caption[]{{
Lowest states using the composite link $V[U]$ and Polyakov loops
in YM-theory at $\beta=9$ with spatial lattice size $L$.
}}
\label{mvl}
\end{figure}
%\vspace*{-0.5cm}
%
Note that this does {\it not} imply an asymptotically free colored state.
The temporal links in the operator
indicate the presence of static charges, in analogy to the Wilson loop.
Here the zero momentum projection switches
off the self-energy of the sources, which thus remain classical fields and
do not affect the energy eigenvalues.

First numerical tests are performed in the 3d theory,
which displays the same confinement properties
but is much more easily accessible numerically.
It is expedient to first test this new operator in a Higgs model,
which in its broken phase has physical states with the quantum numbers of
the gluon, the W-bosons, with detailed results available \cite{us}. 
\fig \ref{comp} compares the W-mass measured in these works by the standard
operator $V[U,\phi]=\tr\left(T^a\phi^\dag(x)U(x)\phi(x+\hat{\mu})\right)$, with
results obtained from the composite links $V[U]$. Full agreement is observed
for different lattice spacings, which also implies that the energies \eq (\ref{finalen})
do have a continuum limit. On the other hand, when simulated in the pure gauge
theory, the corresponding masses grow linearly with the box size, as shown in 
\fig \ref{mvl}.
This effect stems from the non-locality of the gluon operator, which
depends on all link variables in a t-slice. On a periodic torus and in a
confining regime, it will thus
project predominantly on torelonic states. This is evinced by correlations of
the traceless part of
spatial Polyakov loops $L[V]$ constructed from the composite links, which produce
identical results. The same observation is made for the $0^{++}$ operator
$\tr V_i(x)$, which almost exclusively couples to the torelon, even when the 
latter becomes heavier than the lightest scalar glueball. 
In a confining dynamics, \eq (\ref{ofinal}) thus appears to only pick up torelonic states
and is inconclusive regarding the mass scale associated with the gluon. 

\section{THE GLUON PROPAGATOR AND STATIC MESONS}

It is then necessary to construct an alternative operator coupling to the
states we are interested in. This can be achieved by 
a limit procedure. Adding the scalar action
\be
S_{\phi}[U,\phi]=\sum_x\left\{-|D_\mu\phi(x)|^2 + m_0^2|\phi(x)|^2\right\}
\ee
to the pure gauge theory results in QCD with scalar quarks. 
In the limit $m_0\rightarrow \infty$ the scalars become static sources propagating 
in time only,
their propagator being known exactly to consist of
temporal Wilson lines.  

Scalar and vector mesons are described by
$S(x)=\phi^\dag(x)\phi(x),V(x)={\rm Im}(\phi^\dag(x) D_i(x)\phi(x))$, respectively.
In the static limit the scalar fields do not contribute to angular
momentum, nor can they be excited into higher quantum states since they are quenched.
Consequently the mass difference $m_A\equiv M_V-M_S$ is a pure gauge quantity in that
limit, characterizing an excitation with the quantum numbers of the gluon.
Moreover, integrating the scalars out analytically,
one obtains for the ratio of correlators 
$\langle V(x)V(y)\rangle_c/\langle S(x)S(y)\rangle_c \sim$
\be
%\langle V(x)V(y)\rangle_c/\langle S(x)S(y)\rangle_c\sim\\
\frac{\int DA \;\tr\left(A_i(x)U_0(x,y)A_i(y)U^\dag_0(x,y)\right)
\ex^{-S_{YM}} }
{\int DA \; \tr\left(U_0(x,y)
U^\dag_0(x,y)\right)\ex^{-S_{YM}}}.
\ee
In temporal gauge this reduces to the gluon propagator again,
which then relates the mass difference $M_V-M_S$ to the pole mass of the gluon.

\begin{figure}[ht]
\centerline{\epsfxsize=7.0cm\hspace*{0cm}\epsfbox{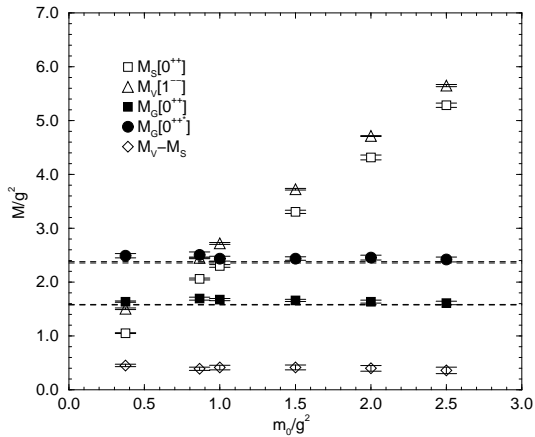}}
\vspace*{-1cm}
\caption[a]{{
The lowest states in 3d scalar QCD. $M_G$ denotes scalar glueballs, $M_{S,V}$
scalar and vector mesons.}}
\label{mhiggs}
\end{figure}

Numerical results, again for 2+1 dimensional SU(2), are shown in \fig \ref{mhiggs}.
With increasing scalar mass, the measured glueball states $M_G$ attain their pure
gauge values indicated by the dashed lines. The scalar bound states 
move out of the spectrum, with $M_V-M_S$ approximately constant. At the largest
scalar mass one finds $m_A=0.37(6)g^2$, or $M_G/m_A\approx 4.2$.

These considerations are readily extended to QCD and heavy quark physics.
Averaging over quark spins, one finds the splittings $\Delta_{1P-1S}(\bar{c}c)=418.5$ MeV,
$\Delta_{1P-1S}(\bar{b}b)=416$ MeV \cite{prev} to be nearly equal, despite a factor of three
difference in the charm and bottom quark mass. According to the above, this splitting
is determined by the QCD gluon pole mass, up to finite quark mass corrections.
With the lightest pure SU(3) glueball at $\sim 1700$ MeV, one has indeed $M_G/m_A\approx 4.1$,
in striking analogy to the 3d SU(2) model.
A calculation of the static limits is in progress.

\end{document}